\pdfoutput=1
\documentclass{article}
\usepackage[preprint]{spconfa4}
\usepackage{amsmath,graphicx}
\usepackage[style=ieee,giveninits=true,url=false,doi=false,isbn=false,maxbibnames=3]{biblatex}
\AtEveryBibitem{\clearlist{language}}
\AtEveryBibitem{\clearfield{eprint}}
\usepackage{glossaries}
\usepackage{bbold}
\usepackage{empheq}
\usepackage[subtle,tracking=normal]{savetrees}
\usepackage{booktabs}
\usepackage{tabularx}
\usepackage[hidelinks]{hyperref}
\usepackage{svg}
\usepackage{siunitx}

\glsdisablehyper
\title{Deep Multi-Frame MVDR Filtering For Binaural Noise Reduction}
\name{Marvin Tammen, Simon Doclo\thanks{This work was funded by the Deutsche Forschungsgemeinschaft (DFG, German Research Foundation) under Germany's Excellence Strategy - EXC 2177/1 - Project ID 390895286.}}
\address{Department of Medical Physics and Acoustics and Cluster of Excellence Hearing4all\\
	University of Oldenburg, Germany\\$\lbrace$marvin.tammen, simon.doclo$\rbrace$@uni-oldenburg.de}
\newcommand*{\bPhi}[2][]{\ensuremath{\boldsymbol{\Phi}^{#1}_{#2,t}}}

\newcommand*{\bgamma}[2][]{\ensuremath{\boldsymbol{\gamma}{}^{#1}_{#2,m,t}}}

\newcommand*{\tran}{\ensuremath{{}^{\mkern-1.5mu\mathsf{T}}}}

\newcommand*{\hermconj}{\ensuremath{{}^{\mathsf{H}}}}

\newcommand{\mt}{\ensuremath{_{m,t}}}

\DeclareMathOperator*{\argmin}{\ensuremath{\mathrm{argmin}}}

\newacronym{stft}{STFT}{short-time Fourier transform}
\newacronym{mvdr}{MVDR}{minimum variance distortionless response}
\newacronym{mfmvdr}{MFMVDR}{multi-frame \gls{mvdr}}
\newacronym{ifc}{IFC}{interframe correlation}
\newacronym{dnn}{DNN}{deep neural network}
\newacronym{dns}{DNS}{deep noise suppression}
\newacronym{mos}{MOS}{mean opinion score}
\newacronym{tcn}{TCN}{temporal convolutional network}
\newacronym{pesq}{PESQ}{perceptual evaluation of speech quality}
\newacronym{stoi}{STOI}{short-time objective intelligibility}
\newacronym{fwssnr}{FWSSNR}{frequency-weighted segmental \gls{snr}}
\newacronym{rtf}{RTF}{real-time factor}
\newacronym{snr}{SNR}{signal-to-noise-ratio}
\newacronym{sir}{SIR}{signal-to-interference-ratio}
\newacronym{sisdr}{SI-SDR}{scale-invariant signal-to-distortion-ratio}
\newacronym{fir}{FIR}{finite impulse response}
\newacronym{stcv}{STCV}{spatio-temporal correlation vector}
\newacronym{stcm}{STCM}{spatio-temporal covariance matrix}
\newacronym{msae}{MSAE}{mean spectral absolute error}
\newacronym{cec1}{CEC1}{first Clarity Enhancement Challenge}
\newacronym{dns3}{DNS3}{third Deep Noise Suppression Challenge}
\newacronym{brir}{BRIR}{binaural room impulse response}

\begin{document}
\ninept
\maketitle
\begin{abstract}
To improve speech intelligibility and speech quality in noisy environments, binaural noise reduction algorithms for head-mounted assistive listening devices are of crucial importance.
Several binaural noise reduction algorithms such as the well-known binaural \gls{mvdr} beamformer have been proposed, which exploit spatial correlations of both the target speech and the noise components.
Furthermore, for single-microphone scenarios, multi-frame algorithms such as the \gls{mfmvdr} filter have been proposed, which exploit temporal instead of spatial correlations.
In this contribution, we propose a binaural extension of the \gls{mfmvdr} filter, which exploits \emph{both} spatial and temporal correlations.
The binaural \gls{mfmvdr} filters are embedded in an end-to-end deep learning framework, where the required parameters, i.e., the speech spatio-temporal correlation vectors as well as the (inverse) noise spatio-temporal covariance matrix, are estimated by \glspl{tcn} that are trained by minimizing the mean spectral absolute error loss function.
Simulation results comprising measured binaural room impulses and diverse noise sources at signal-to-noise ratios from -5\,$\mathrm{dB}$ to 20\,$\mathrm{dB}$ demonstrate the advantage of utilizing the binaural \gls{mfmvdr} filter structure over directly estimating the binaural multi-frame filter coefficients with \glspl{tcn}.
\end{abstract}
\glsresetall
\section{Introduction}
\label{sec: Introduction}
In many speech communication scenarios, head-mounted assistive listening devices such as binaural hearing aids capture not only the target speaker, but also ambient noise, resulting in a degradation of speech quality and speech intelligibility.
Hence, several binaural noise reduction algorithms have been proposed, which typically assume that adjacent \gls{stft} coefficients are uncorrelated over time.
This assumption is suitable when considering sufficiently long frames and a small frame overlap.
In that case, the speech \gls{stft} coefficients at a left and right reference microphone can be estimated by applying (complex-valued) single-frame binaural filters to the available microphone signals. 
Several approaches have been proposed to estimate these single-frame binaural filters, which can be categorized into statistical model-based approaches (e.g., ~\cite{doclo_2015_multichannel,hadad_2016_binaural,gannot_2017_consolidated,doclo_2018_binaural}) and supervised learning-based approaches (e.g., \cite{moore_2018_binauralb,sun_2019_deepb,han_2020_real,sun_2020_supervised,kim_2021_mimo,borgstrom_2021_speaker,green_2022_speech}).
While the statistical model-based approaches can be mainly differentiated w.r.t. their underlying optimization problem and how the required parameters are estimated, 
the supervised learning-based approaches mainly differ in the used \gls{dnn} architecture and loss function. 

With the goal of exploiting temporal correlations between neighboring \gls{stft} coefficients, multi-frame methods have been proposed for both single- and multi-microphone noise reduction, which apply (complex-valued) multi-frame filters to the most recent noisy \gls{stft} coefficients of each microphone.
Similarly to the single-frame methods mentioned above, several approaches have been proposed to estimate these multi-frame filters, which can again be categorized into statistical model-based approaches (e.g., \cite{huang_2012_multi,habets_2012_multi}) and supervised learning-based approaches (e.g., \cite{mack_2020_deep,aroudi_2020_cognitive,tammen_2021_deep,zhang_2021_multi,wang_2021_sequential}). 
In contrast to the single-frame approaches, however, there is a lack of studies that considered multi-frame approaches for \emph{binaural} noise reduction.

\begin{figure*}[tb]
	\centering
	\includegraphics[width=0.9\linewidth]{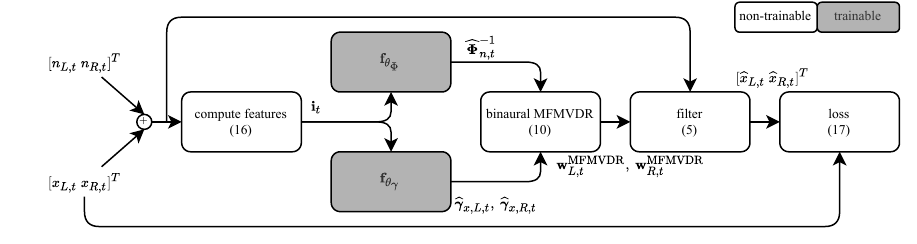}
	\caption{Block diagram of the proposed deep binaural \gls{mfmvdr} filter.}
	\label{fig: block diagram}
\end{figure*}

Aiming at utilizing both spatial correlations as in the binaural \gls{mvdr} beamformer~\cite{doclo_2015_multichannel,gannot_2017_consolidated} and temporal correlations as in the \gls{mfmvdr} filter~\cite{huang_2012_multi,tammen_2021_deep}, we propose to extend the \gls{mfmvdr} filter to binaural listening scenarios.
To implement the binaural \gls{mfmvdr} filter, estimates of the speech \glspl{stcv} as well as the (inverse) noise \gls{stcm} are required.
Similarly as in \cite{tammen_2021_deep}, the binaural \gls{mfmvdr} filter is embedded in an end-to-end supervised learning framework as shown in Fig. \ref{fig: block diagram}, where all required parameters are estimated using \glspl{tcn} that are trained using the \gls{msae} loss function~\cite{wang_2020_complex}.
Simulation results using measured binaural room impulse responses from \cite{kayser_2009_database} as well as clean speech and noise from the \gls{dns3}~\cite{reddy_2021_interspeechb} at \glspl{snr} from \SIrange{-5}{20}{\decibel} show that the proposed deep binaural \gls{mfmvdr} filter outperforms directly estimating the single- or multi-frame binaural filter coefficients using \glspl{tcn}, i.e., without exploiting the structure of the deep binaural \gls{mfmvdr} filter.

\section{Signal Model}
\label{sec: Signal Model}
\glslocalreset{stcv}
\glslocalreset{stcm}
We consider an acoustic scenario with a single speech source and a single noise source, both located in a reverberant room, recorded by binaural hearing aids with $M$ microphones.
In the \gls{stft} domain, the noisy microphone signals $y_{m,f,t}$ are given by
\begin{equation}\label{eq: signal model}
	y_{m,f,t} = x_{m,f,t} + n_{m,f,t},
\end{equation}
where $x_{m,f,t}$ and $n_{m,f,t}$ denote the speech and noise components, respectively, at the $m$-th microphone, the $f$-th frequency bin, and the $t$-th time frame. 
Since all frequency bins are processed independently, the index $f$ will be omitted in the remainder of this paper.

In \emph{single}-microphone multi-frame noise reduction algorithms~\cite{huang_2012_multi,tammen_2021_deep}, the noisy multi-frame vector $\bar{\mathbf{y}}\mt \in \mathbb{C}^N$ is defined as
\begin{equation}\label{eq: noisy multi-frame vector}
	\bar{\mathbf{y}}\mt = \begin{bmatrix}	y\mt & \dots & y_{m,t-N+1} \end{bmatrix}\tran,
\end{equation}
with $\circ\tran$ denoting the transpose operator, such that \eqref{eq: signal model} can be written as $\bar{\mathbf{y}}\mt = \bar{\mathbf{x}}\mt + \bar{\mathbf{n}}\mt$.
In this case, using a complex-valued multi-frame filter $\bar{\mathbf{w}}\mt \in \mathbb{C}^N$, the speech component $x\mt$ is estimated as
\begin{equation}\label{eq: speech estimate multi-frame}
	\widehat{x}\mt = \bar{\mathbf{w}}\hermconj\mt \bar{\mathbf{y}}\mt,
\end{equation}
where $\circ\hermconj$ denotes the conjugate transpose operator.

In \emph{multi}-microphone multi-frame noise reduction algorithms~\cite{habets_2012_multi,zhang_2021_multi,wang_2021_sequential}, the noisy multi-microphone multi-frame vector $\mathbf{y}\mt \in \mathbb{C}^{N M}$ is defined as
\begin{equation}
	\mathbf{y}_t = \begin{bmatrix} \bar{\mathbf{y}}\tran_{1,t} & \dots & \bar{\mathbf{y}}\tran_{M,t} \end{bmatrix}\tran,
\end{equation}
such that \eqref{eq: signal model} can be written as $\mathbf{y}_t = \mathbf{x}_t + \mathbf{n}_t$.
Without loss of generality, in this paper we consider the case $M=2$, with one hearing aid per side and one microphone per hearing aid, i.e., $m \in \{L,R\}$, where $L$ and $R$ denote the left and right side, respectively.
In this case, using (complex-valued) binaural multi-frame filters $\mathbf{w}\mt \in \mathbb{C}^{2N}$ with $2N$ taps each, the binaural speech components are estimated as
\begin{equation}\label{eq: binaural speech estimate multi-frame}
	\widehat{x}\mt = \mathbf{w}\hermconj\mt \mathbf{y}_t.
\end{equation}
Assuming that the speech and noise components are spatio-temporally uncorrelated, the noisy \gls{stcm} $\bPhi{y} = \mathcal{E} \{ \mathbf{y}_t \mathbf{y}\hermconj_t \} \in \mathbb{C}^{2N \times 2N}$, with $\mathcal{E}\{\circ\}$ the expectation operator, can be written as
\begin{equation}\label{eq: noisy stcm}
	\bPhi{y} =  \bPhi{x} + \bPhi{n},
\end{equation}
where $\bPhi{x}$ and $\bPhi{n}$ are defined similarly as $\bPhi{y}$.

In order to exploit speech correlations across successive time frames, it has been proposed in \cite{huang_2012_multi} to decompose the (single-microphone) multi-frame speech vector into a temporally correlated and a temporally uncorrelated component.
Similarly, the binaural multi-frame speech vector $\mathbf{x}_t$ can be decomposed into a spatio-temporally correlated and a spatio-temporally uncorrelated component w.r.t. the current left or the right speech \gls{stft} coefficient $x_{m,t}$:
\begin{align}\label{eq: speech vector decomposition}
	\mathbf{x}_t &= \underbrace{\bgamma{x} x_{m,t}}_{\text{correlated}} + \underbrace{\mathbf{x}'_{m,t}}_{\text{uncorrelated}}
\end{align}
The highly time-varying left or right speech \gls{stcv} $\bgamma{x} \in \mathbb{C}^{2N}$ describes the correlation between the $N$ most recent left and right speech \gls{stft} coefficients and the current left or the right speech \gls{stft} coefficient $x_{m,t}$, and it is defined as
\begin{equation}\label{eq: speech stcv}
	\bgamma{x} = \frac{\mathcal{E} \{ \mathbf{x}_t x^*_{m,t} \}}{\mathcal{E} \{ |x_{m,t}|^2 \}},
\end{equation}
where $\circ^*$ denotes the conjugate operator and with $\mathbf{e}\tran_L \boldsymbol{\gamma}_{x,L,t} = \mathbf{e}\tran_R \boldsymbol{\gamma}_{x,R,t} = 1$.
Here, $\mathbf{e}_L$ and $\mathbf{e}_R$ denote selection vectors with their first or $N+1$-th element equal to \num{1}, respectively, and the other elements equal to \num{0}.

\section{Deep Binaural Multi-Frame MVDR Filter}
\label{sec: Deep Binaural Multi-Frame MVDR Filter}
Aiming at minimizing the output noise power spectral density while leaving the correlated speech component undistorted, in \cite{huang_2012_multi} the \gls{mfmvdr} filter for single-microphone noise reduction has been proposed.
In this paper, we propose to extend the single-microphone \gls{mfmvdr} filter to binaural scenarios by considering the spatio-temporal correlations of the speech and noise components for the left and right side, i.e.,
\begin{equation}\label{eq: optimization problem}
	\argmin_{\mathbf{w}\mt} \quad \mathbf{w}\hermconj\mt \bPhi{n} \mathbf{w}\mt \quad \text{s.t.} \quad \mathbf{w}\hermconj\mt \bgamma{x} = 1.
\end{equation}
Solving this optimization problem, the binaural \gls{mfmvdr} filters are given by
\begin{empheq}[box=\fbox]{align}\label{eq: binaural mfmvdr filter}
	\mathbf{w}^{\mathrm{MFMVDR}}\mt = \frac{\bPhi[-1]{n} \bgamma{x}}{\boldsymbol{\gamma}\hermconj_{x,m,t} \bPhi[-1]{n} \bgamma{x}}
\end{empheq}
As has been shown for the \emph{single-microphone} \gls{mfmvdr} filter~\cite{fischer_2017_sensitivity}, the performance of the (binaural) \gls{mfmvdr} filter depends on how well the required parameters, i.e., the inverse noise \gls{stcm} as well as the speech \glspl{stcv}, are estimated from the noisy \gls{stft} coefficients.
In contrast to using statistical model-based estimators similar to \cite{schasse_2014_estimation}, we embed the binaural \gls{mfmvdr} filter in an end-to-end supervised learning framework similar to \cite{tammen_2021_deep}, with the parameters estimated by \glspl{tcn} (see Fig. \ref{fig: block diagram}).
The \glspl{tcn} are trained by minimizing the \gls{msae} loss function~\cite{wang_2020_complex} computed at the output of the deep binaural \gls{mfmvdr} filter instead of providing explicit parameter labels.
A-priori knowledge about the properties of the estimated parameters is exploited as described in the following two sections. 

\subsection{Speech Spatio-Temporal Correlation Vector}
\label{ssec: Speech Spatio-Temporal Correlation Vector}
The left and right speech \glspl{stcv} each are two $2N$-dimensional complex-valued vectors (cf. \eqref{eq: speech stcv}), hence consisting of $8N$ \emph{real}-valued coefficients $\mathbf{h}^{\mathbb{R}}_{\gamma,t} \in \mathbb{R}^{8N}$ ($4N$ for the real part and $4N$ for the imaginary part).
To estimate these real-valued coefficients, we propose to use a \gls{tcn} $\mathbf{f}_{\theta_\gamma}$ with parameters $\boldsymbol{\theta}_\gamma$, which is fed input features $\mathbf{i}_{t}$ derived from the noisy \gls{stft} coefficients, i.e.,
\begin{equation}\label{eq: dnn output speech stcv}
	\widehat{\mathbf{h}}^{\mathbb{R}}_{\gamma,t} = \mathbf{f}_{\theta_\gamma} \{ \mathbf{i}_{t} \},
\end{equation}
with the features $\mathbf{i}_{t}$ defined in \eqref{eq: features}.
To construct a $4N$-dimensional complex-valued vector $\widehat{\mathbf{h}}^{\mathbb{C}}_{\gamma,t}$ from the $8N$-dimensional real-valued vector $\widehat{\mathbf{h}}^{\mathbb{R}}_{\gamma,t}$, the first $4N$ elements of $\widehat{\mathbf{h}}^{\mathbb{R}}_{\gamma,t}$ are used for the real components and the second $4N$ elements are used for the imaginary components, i.e.,
\begin{equation}
	\widehat{\mathbf{h}}^{\mathbb{C}}_{\gamma,t} = [\widehat{\mathbf{h}}^{\mathbb{R}}_{\gamma,t}]_{0:4N-1} + j \ [\widehat{\mathbf{h}}^{\mathbb{R}}_{\gamma,t}]_{4N:8N-1} ,
\end{equation}
where $j^2 = -1$.
To ensure that the first or $N+1$-th element of the speech \glspl{stcv} is equal to $1$ (cf. \eqref{eq: speech stcv}), the speech \glspl{stcv} are finally obtained as
\begin{equation}\label{eq: speech stcv estimate}
	\widehat{\boldsymbol{\gamma}}_{x,L,t} = \frac{[\widehat{\mathbf{h}}^{\mathbb{C}}_{\gamma,t}]_{0:2N-1}}{\mathbf{e}\tran_L [\widehat{\mathbf{h}}^{\mathbb{C}}_{\gamma,t}]_{0:2N-1}}, \quad \widehat{\boldsymbol{\gamma}}_{x,R,t} = \frac{[\widehat{\mathbf{h}}^{\mathbb{C}}_{\gamma,t}]_{2N:4N-1}}{\mathbf{e}\tran_R [\widehat{\mathbf{h}}^{\mathbb{C}}_{\gamma,t}]_{2N:4N-1}}.
\end{equation}

\subsection{Spatio-Temporal Covariance Matrices}
\label{ssec: Spatio-Temporal Covariance Matrices}
Since the $2N \times 2N$-dimensional \gls{stcm} $\bPhi{n}$ can be assumed to be Hermitian positive-definite, also its inverse $\bPhi[-1]{n}$ as required in \eqref{eq: binaural mfmvdr filter} can be assumed to be Hermitian positive-definite.
Hence, $\bPhi[-1]{n}$ has a unique Cholesky decomposition~\cite{cholesky_1924_note}:
\begin{equation}\label{eq: cholesky decomposition}
	\bPhi[-1]{n} = \mathbf{L}_t \mathbf{L}\hermconj_t,
\end{equation}
with $\mathbf{L}_t \in \mathbb{C}^{2N \times 2N}$ a lower triangular matrix with positive real-valued diagonal.
Due to its structure, $\mathbf{L}$ is determined by $(2N)^2$ real-valued coefficients.
Similarly to the procedure for estimating the speech \glspl{stcv}, we use a \gls{tcn} $\mathbf{f}_{\theta_\Phi}$ with parameters $\boldsymbol{\theta}_\Phi$, which is fed input features $\mathbf{i}_{t}$, to estimate these real-valued coefficients $\widehat{\mathbf{h}}^{\mathbb{R}}_{\Phi,t} \in \mathbb{R}^{(2N)^2}$, i.e.,
\begin{equation}\label{eq: dnn output noise stcm}
	\widehat{\mathbf{h}}^{\mathbb{R}}_{\Phi,t} = \mathbf{f}_{\theta_\Phi} \{ \mathbf{i}_{t} \}.
\end{equation}
Using $\widehat{\mathbf{h}}^{\mathbb{R}}_{\Phi,t}$, the lower triangular matrix with positive real-valued diagonal $\widehat{\mathbf{L}}_t$ is assembled.
Finally, an estimate of $\bPhi[-1]{n}$ is obtained using \eqref{eq: cholesky decomposition} by replacing $\mathbf{L}_t$ with its estimate $\widehat{\mathbf{L}}_t$.

\section{Simulations}
\label{sec: Simulations}
In this section, the binaural noise reduction performance of the proposed deep binaural \gls{mfmvdr} filter is compared with a number of baseline algorithms, which are described in Section \ref{ssec: Baseline Algorithms}.
Sections \ref{ssec: Dataset} and \ref{ssec: Settings} deal with the used datasets and the simulation settings, respectively.
In Section \ref{ssec: Results}, the simulation results are presented in terms of the \gls{pesq}~\cite{rix_2001_perceptual} and \gls{fwssnr}~\cite{hu_2008_evaluation} improvement.

\subsection{Baseline Algorithms}
\label{ssec: Baseline Algorithms}
The following baseline algorithms have been considered to allow investigating the effect of not using vs. using the proposed deep binaural \gls{mfmvdr} structure for binaural multi-frame filtering.
To achieve this goal, for the baseline algorithms the binaural multi-frame filters in \eqref{eq: binaural speech estimate multi-frame} are not obtained using the binaural \gls{mfmvdr} structure.
Instead, the real and imaginary components of the baseline binaural multi-frame filters are directly estimated by a \gls{tcn}, i.e., without the intermediate steps of speech \glspl{stcv} and inverse noise \gls{stcm} estimation and computation of \eqref{eq: binaural mfmvdr filter}.
In addition, we investigate the effect of binaural single-frame vs. binaural multi-frame filtering.
More specifically, we use the following end-to-end supervised learning-based baseline algorithms:
\begin{description}
	\item[direct binaural single-frame filtering] With $N=1$ and $\mathbf{w}^{\mathrm{B1}}\mt \in \mathbb{C}^{2}$, only spatial filtering is performed.
	The filter coefficients are estimated using a \gls{tcn} $\mathbf{f}_{\mathrm{B1}}$ with parameters $\boldsymbol{\theta}_{\mathrm{B1}}$, i.e., $\mathbf{w}^{\mathrm{B1}}\mt = \mathbf{f}_{\mathrm{B1}} \{ \mathbf{i}_{t} \}$.
	The real and imaginary parts of the filter coefficients $\mathbf{w}^{\mathrm{B1}}\mt$ are bounded to $[-1, 1]$ using a hyperbolic tangent activation function.
	\item[direct binaural multi-frame filtering] With $N=3$ and $\mathbf{w}^{\mathrm{B2}}\mt \in \mathbb{C}^{2N}$, both spatial and temporal filtering are performed.
	The filter coefficients are estimated using a \gls{tcn} $\mathbf{f}_{\mathrm{B2}}$ with parameters $\boldsymbol{\theta}_{\mathrm{B2}}$, i.e., $\mathbf{w}^{\mathrm{B2}}\mt = \mathbf{f}_{\mathrm{B2}} \{ \mathbf{i}_{t} \}$.
	The real and imaginary parts of the filter coefficients $\mathbf{w}^{\mathrm{B2}}\mt$ are bounded to $[-1, 1]$ using a hyperbolic tangent activation function.
	These bounds are motivated by \cite{mack_2020_deep}.
\end{description}

\subsection{Dataset}
\label{ssec: Dataset}
\glslocalreset{dns3}
To train and validate the considered algorithms, we used simulated \glspl{brir} from the training subset of the \gls{cec1} dataset~\cite{graetzer_2021_clarity} as well as clean speech (English read book sentences) and noise from the training subset of the \gls{dns3} dataset~\cite{reddy_2021_interspeechb}.
These \glspl{brir} were simulated by considering a randomly positioned directed speech source and an omnidirectional noise point source captured by binaural behind-the-ear hearing aids in randomly sized rooms with "low to moderate" reverberation, i.e., around \SIrange{0.2}{0.4}{\second}.
The speech source was always located at an angle within $\ang{\pm 30}$ w.r.t. the listener, while the noise source could be positioned everywhere in the room except for less than \SI{1}{\metre} from the walls or the listener.
Surface absorption coefficients were varied to simulate various room characteristics such as doors, windows, curtains, rugs, or furniture.
In total, 6000 room configurations were considered.
Clean speech and noise were convolved with their corresponding \glspl{brir} before being mixed at better ear \glspl{snr} from \SIrange{0}{15}{\decibel}.
In total, the training and validation datasets have a length of \SI{80}{h} and \SI{20}{h}, respectively. 

To evaluate the considered algorithms, we used measured \glspl{brir} from the dataset proposed in \cite{kayser_2009_database} as well as clean speech and noise from the official test subset of the \gls{dns} dataset~\cite{reddy_2020_interspeechb}.
The dataset in \cite{kayser_2009_database} comprises \glspl{brir} measured with binaural behind-the-ear hearing aids ``for multiple, realistic head and sound-source positions in four natural environments reflecting daily-life communication situations with different reverberation times''.
The configuration of these hearing aids matches the configuration considered in the training and validation datasets.
Clean speech and noise were convolved with the \glspl{brir} before being mixed at better ear \glspl{snr} from \SIrange{-5}{20}{\decibel}.
In total, 100 utterances, each of length \SI{10}{s}, were considered in the evaluation.
Note that, especially due to the use of simulated vs. measured \glspl{brir}, there is considerable mismatch between the training and validation datasets on the one hand and the evaluation dataset on the other hand.
All datasets were used at a sampling frequency of \SI{16}{kHz}.

\subsection{Settings}
\label{ssec: Settings}
For the \gls{stft} used in all considered algorithms, $\sqrt{\text{Hann}}$ windows with a frame length of \SI{8}{\milli\second} and a frame shift of \SI{2}{\milli\second} were used for both analysis and synthesis.
As input features, we used a concatenation of the logarithmic magnitude, the cosine of the phase, and the sine of the phase, of the noisy left and right \gls{stft} coefficients, i.e.,
\begin{align}\label{eq: features}
	\mathbf{i}_{m,t} &=&& \begin{bmatrix}
		\log_{10} |y_{m,t}| & \cos (\angle y_{m,t}) & \sin (\angle y_{m,t})
	\end{bmatrix}\tran\nonumber\\
	\mathbf{i}_{t} &=&& \begin{bmatrix}
		\mathbf{i}\tran_{L,t} & \mathbf{i}\tran_{R,t}
	\end{bmatrix}]\tran,
\end{align}
where $\angle \circ$ denotes the phase of $\circ$.
Note that both the cosine and sine of the noisy phase are chosen to prevent an ambiguous phase representation.

The multi-frame algorithms use $N=5$ frames, resulting in the capability of exploiting temporal correlations within \SI{16}{\milli\second}.
To decrease distortion of the speech and residual noise components, a minimum gain of \SI{-20}{\decibel} was included in all algorithms.

To estimate the required parameters of the deep binaural \gls{mfmvdr} filter or the filter coefficients of the baseline algorithms, we used \glspl{tcn}, with their hyperparameters fixed to 2 stacks of 6 layers, yielding a temporal receptive field size of \SI{512}{\milli\second}.
Since the deep binaural \gls{mfmvdr} filter uses two \glspl{tcn} and the number of real-valued coefficients differs per considered algorithm, the hidden dimension size of the \glspl{tcn} was varied per algorithm to result in similar numbers of trainable weights for all algorithms, i.e., \num{6.2e6}.
While also the other hyperparameters could have been varied to this end, only varying the hidden dimension size results in \glspl{tcn} with the same temporal receptive field size, which is required for a fair comparison.
To prevent division by \num{0}, a small constant was added to the denominator in \eqref{eq: speech stcv estimate}.

As loss function, the \gls{msae} proposed in \cite{wang_2020_complex} was used, where the loss was averaged across the batch, the left and right output signals, and the frequency bins and time frames, i.e.,
\begin{align}\label{eq: loss}
	L_{b,m,f,t} &= \beta \left|x_{b,m,f,t} - \widehat{x}_{b,m,f,t} \right| + (1 - \beta) \left||x_{b,m,f,t}| - |\widehat{x}_{b,m,f,t}| \right|\nonumber\\
	L &= \frac{1}{2 B F T} \sum_{b=0}^{B-1} \sum_{m \in \{ L,R \}} \sum_{f=0}^{F-1} \sum_{t=0}^{T-1} L_{b,m,f,t},
\end{align}
where $B$ denotes the batch size, $F$ and $T$ denote the numbers of frequency bins and time frames in an utterance, and $\beta=0.4$~\cite{wang_2020_complex}.

The \glspl{tcn} were implemented based on the official Conv-TasNet implementation\footnote{\url{https://github.com/naplab/Conv-TasNet}}, and they were trained for a maximum of \num{150} epochs with early stopping using the AdamW optimizer~\cite{loshchilov_2019_decoupled}.
The learning rate was initialized as \num{3e-4}, and it was halved after \num{3} epochs without an improvement on the validation dataset.
Gradient $\ell_2$-norms were clipped to \num{5}, and the batch size was \num{8}.

The simulations were implemented using PyTorch 1.10~\cite{paszke_2019_pytorch} and performed on NVIDIA GeForce\textregistered RTX A5000 graphics cards.
A PyTorch implementation of the compared algorithms as well as the model weights used in the evaluation will be made publicly available upon publication.

\subsection{Results}
\label{ssec: Results}
\begin{figure}[tb]
	\centering
	\includesvg[width=0.45\textwidth]{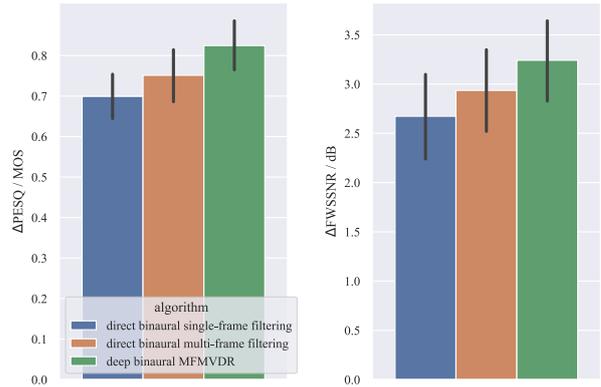}
	\caption{Mean and standard deviation of the \gls{pesq} and \gls{fwssnr} improvements obtained on the evaluation dataset. The mean noisy \gls{pesq} score is \SI{1.74}{MOS} and the mean noisy \gls{fwssnr} score is \SI{14.08}{\decibel}.}
	\label{fig: results}
\end{figure}
For all considered algorithms, Fig.~\ref{fig: results} depicts the improvement in terms of \gls{pesq} and \gls{fwssnr} w.r.t. the noisy microphone signals on the evaluation dataset.
Note that, similarly as for the \gls{msae} loss function in \eqref{eq: loss}, \gls{pesq} and \gls{fwssnr} improvements are simply averaged across the left and right output signals~\cite{borgstrom_2021_speaker}.

First, a considerable improvement in terms of \gls{pesq} and \gls{fwssnr} can be observed for all algorithms, with the deep binaural \gls{mfmvdr} filter outperforming the baseline algorithms.
Second, comparing the baseline algorithms, it can be observed that increasing the degrees of freedom of the filter, i.e., by allowing for a binaural multi-frame vs. a binaural single-frame filter, improves binaural noise reduction performance.
Third, by enforcing the binaural \gls{mfmvdr} structure on the binaural multi-frame filter, binaural noise reduction performance is further increased.

Audio examples for the compared algorithms are available online\footnote{\url{https://uol.de/en/sigproc/research/audio-demos/binaural-noise-reduction/deep-bmfmvdr}}.

\section{Conclusion}
\label{sec: Conclusions}
In this paper we proposed a binaural extension of the \gls{mfmvdr} filter, which is capable of utilizing both spatial and temporal correlations of the speech and noise components.
To estimate the speech \glspl{stcv} as well as the inverse noise \gls{stcm} required by the binaural \gls{mfmvdr} filter, we use \glspl{tcn}, which are trained by embedding the binaural \gls{mfmvdr} filter in an end-to-end supervised learning framework and minimizing the \gls{msae} loss function.
Simulations comprising measured binaural room impulse responses as well as diverse noise sources at \glspl{snr} in \SIrange{-5}{20}{\decibel} demonstrate the advantage of binaural multi-frame filtering over binaural single-frame filtering as well as employing the binaural \gls{mfmvdr} structure over directly estimating the single- or multi-frame binaural filters using \glspl{tcn}.


\printbibliography

\end{document}